\documentclass[sigconf]{acmart}
\usepackage{relsize}
\usepackage{lmodern}
\usepackage[svgnames,table]{xcolor} 
\usepackage{enumitem}
\usepackage{graphicx}
\usepackage{listings}
\usepackage{xcolor}
\AtBeginDocument{%
  }

\acmConference[Conference acronym 'XX]{Make sure to enter the correct
  conference title from your rights confirmation email}{June 03--05,
  2018}{Woodstock, NY}
\acmISBN{978-1-4503-XXXX-X/2018/06}

\usepackage{amsmath,amssymb,amsfonts}
\usepackage{lineno}

\usepackage{array}
\usepackage{booktabs}
\usepackage{siunitx}
\newcolumntype{L}[1]{>{\raggedright\let\newline\\\arraybackslash\hspace{0pt}}m{#1}}
\newcommand{\subrowindent}{\hspace*{1em}}

\usepackage{listings}
\usepackage{upquote} 

\definecolor{codebg}{HTML}{F8F9FB}
\definecolor{codeborder}{HTML}{E5E7EB}
\definecolor{jsonstring}{HTML}{0366D6}
\definecolor{jsonpunct}{HTML}{383A42}
\definecolor{jsondelim}{HTML}{383A42}
\definecolor{jsonconst}{HTML}{D73A49}

\lstdefinelanguage{json}{
  basicstyle=\ttfamily\small,
  showstringspaces=false,
  morestring=[b]",
  stringstyle=\color{jsonstring},
  morecomment=[l]{//},
  morecomment=[s]{/*}{*/},
  literate=
   *{:}{{{\color{jsonpunct}{:}}}}{1}
    {,}{{{\color{jsonpunct}{,}}}}{1}
    {\{}{{{\color{jsondelim}{\{}}}}{1}
    {\}}{{{\color{jsondelim}{\}}}}}{1}
    {[}{{{\color{jsondelim}{[}}}}{1}
    {]}{{{\color{jsondelim}{]}}}}{1}
    {true}{{{\color{jsonconst}{true}}}}{4}
    {false}{{{\color{jsonconst}{false}}}}{5}
    {null}{{{\color{jsonconst}{null}}}}{4},
}

\lstdefinestyle{mintedlike}{
  backgroundcolor=\color{codebg},
  frame=single,
  rulecolor=\color{codeborder},
  framerule=0.4pt,
  numbers=left,
  numberstyle=\tiny\color{gray},
  numbersep=6pt,
  xleftmargin=1.2em,
  xrightmargin=1.2em,
  breaklines=true,
  columns=fullflexible,
  upquote=true,
  basicstyle=\ttfamily\small,
  postbreak=\mbox{\textcolor{black}{$\hookrightarrow$}\space}
}

\usepackage{stfloats}
\usepackage[skip=2pt]{caption}
\begin{document}

\title{ACCeLLiuM: Supervised Fine-Tuning for Automated OpenACC Pragma Generation}


\author{Samyak Jhaveri}
\email{samyaknj@uci.edu}
\affiliation{
\institution{University of California Irvine}
\city{Irvine}
\state{California}
\country{USA}
}
\author{Vanessa Klotzmann}
\email{vklotzma@uci.edu}
\affiliation{
\institution{University of California Irvine}
\city{Irvine}
\state{California}
\country{USA}
}
\author{Crista Lopes}
\email{lopes@uci.edu}
\affiliation{
\institution{University of California Irvine}
\city{Irvine}
\state{California}
\country{USA}
}

\renewcommand{\shortauthors}{Anonymous et al.}

\begin{abstract}
The increasing ubiquity of GPUs is accompanied by the increasing complexity of their hardware and parallel programming frameworks. Directive-based parallel programming standards like OpenACC simplify GPU programming to some extent by abstracting away low-level complexities, but a fair amount of expertise is still required in order to use those directives effectively.

We introduce ACCeLLiuM, two open weights Large Language Models specifically fine-tuned for generating expert OpenACC directives for data-parallel loops, along with the supervised fine-tuning  dataset that was used to train them. The ACCeLLiuM SFT dataset contains 4,033 OpenACC pragma-loop pairs mined from public GitHub C/C++ repositories, with 3,223 pairs for training and 810 for testing. Experimental evaluations show a pronounced performance gap in generating correct OpenACC pragmas between base LLMs and our fine-tuned versions. On the held-out test set, base LLMs fail to consistently generate valid pragmas, whereas LLMs fine-tuned on the ACCeLLiuM dataset generate valid pragmas with the correct directive type for $87\%$ of the data-parallel loops, and exact pragmas--including directives, clauses, clause order, and clause variables--for $50\%$ of the cases. Even when not exact, generated pragmas frequently incorporate the correct clauses in a different order than the ground-truth label, or include additional clauses that enable finer control over parallel execution, data movement, and concurrency, offering practical value beyond strict string-matching. By publicly releasing the code, models, and dataset as ACCeLLiuM we hope to establish a reproducible benchmark for LLM-powered OpenACC pragma generation, and lower the barrier to automated GPU offloading of serially written programs. 

\end{abstract}



\maketitle

\section{Introduction}
\label{sec:introduction}
\begin{figure*}[t] 
  \centering
  \includegraphics[width=\linewidth]{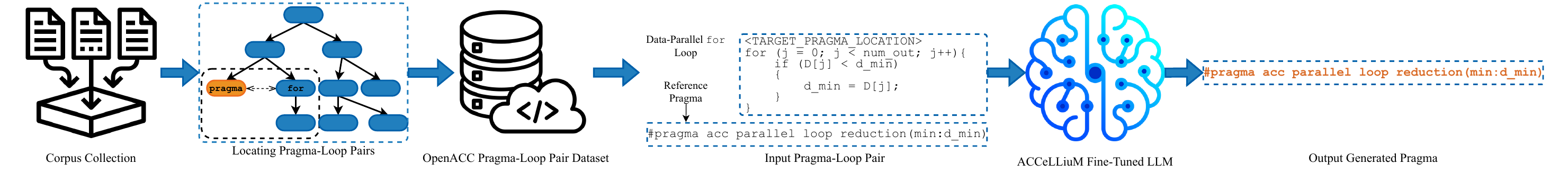}
  \Description{Pipeline: ACCeLLiuM analyzes a nested C loop on the
  left and emits an OpenACC pragma with reduction and data‑movement
  clauses on the right.}
  \caption{ACCeLLiuM detects parallelization patterns in data‑parallel
  loops and automatically generates the appropriate \texttt{OpenACC}
  pragma and clauses.}
  \label{fig:accellium_training_diagram}
\end{figure*}

Graphics Processing Units (GPUs) have become ubiquitous in heterogeneous computing, accelerating computation across a wide range of systems, from exascale High-Performance Computing (HPC) systems to smaller local computing clusters. This widespread adoption is marked by the variety and volume of GPUs from numerous vendors, each with its own distinct architectures and programming frameworks. This growing diversity of GPU hardware is paired with increasing complexity in both the architecture and the proprietary parallel programming frameworks needed to execute parallel programs on these architectures. Although these frameworks are powerful, they require extensive and specialized expertise, making it difficult for scientists and developers to create, optimize, and maintain efficient parallel software at the production level.

CUDA~\cite{nvidia_cuda_programming_guide}, though efficient and popular~\cite{karimi2010performance, fang_performance_comparison_cuda_opencl}, is vendor-locked to NVIDIA hardware. The same applies to ROCm for AMD GPUs. OpenCL, another parallel programming framework, is a portable and vendor-neutral framework that serves as a popular alternative to CUDA and ROCm. However, due to its generalized nature, it abstracts hardware, leading to subpar performance and variability across vendors. With its lower-level API and the need for manual optimization, it can have a steep learning curve and a less user-friendly development experience~\cite{automatic_parallelism_data_dependency}. Obtaining significant efficiency from a GPU using these frameworks requires extensive expertise in writing parallel kernels~\cite{programming_massively_parallel_processors_book}. 

To address these challenges, directive-based programming models like OpenACC~\cite{openacc_specification_2021, parallel_programming_with_openacc_book} have emerged, offering a high-level, more portable approach to offloading compute-intensive workloads. By inserting \texttt{\#pragma} compiler directives, programmers can instruct the compiler to parallelize code by automatically mapping it to the GPU hardware with minimal user intervention. However, avoiding the direct use of these frameworks via high-level compiler directives is not enough to write effective and efficient parallel programs. The manual process of identifying parallelizable loops and, more critically, crafting the correct set of directives and clauses for each loop parallelization and data movement operation between the CPU-GPU system remains a formidable, time-consuming, and error-prone task. It involves an expert understanding of the intricate data dependencies and memory access patterns of complex operations, as well as knowledge of the correct usage of specific clauses to map those operations to parallel execution threads on the GPU hardware~\cite{automatic_parallelism_data_dependency, parallel_programming_with_openacc_book}. Parallelization is inherently challenging, and even experienced developers spend considerable time finding parallelism opportunities and writing parallel versions of sequential programs~\cite{mahmud_autoparllm_2025,unveiling_parallelism_in_serial_code,Li2013DiscoveryOP}. The difficulty of these challenges increases as the scale of the codebase and the complexity of its algorithms increase, as seen in many scientific HPC applications. 

Existing solutions to automatically write OpenACC pragmas have proven less than ideal. Early approaches based on static analysis and compiler tools, such as DawnCC\cite{dawncc_2017} and KernelGen\cite{Mikushin2014KernelGenT}, struggle with the complexities of real-world code, often producing suboptimal pragmas or pragmas for only a specific set of programs. Their functionality depends on unmaintained, highly specialized compilers or compiler modifications. 

Recent work has shown the promise of Large Language Models (LLMs) for generating OpenMP directives to parallelize serial programs using the HPCorpus dataset~\cite{hpcorpus_kadosh, monocoder_kadosh}. However, to the best of our knowledge, there are no publicly available specialized datasets for training LLMs for generating OpenACC pragmas. 

To address this gap, we introduce ACCeLLiuM (Figure~\ref{fig:accellium_training_diagram}) , an open-source bundle that acts as an end-to-end resource for research and practice, comprising (1) a curated dataset of $\approx$ 4,000 OpenACC pragma-loop pairs mined from public GitHub C and C++ repositories, (2) a family of locally deployable open-weights LLMs (built on top of Llama 3.1 70B and CodeLlama 34B) fine-tuned using supervised fine-tuning on this dataset, and (3) an open-source pipeline for dataset creation, model fine-tuning, and evaluation. The objective of fine-tuning the LLMs using our dataset is to train them to annotate data-parallel loops with the correct OpenACC pragmas. Our goal in this study is to explore the feasibility and benefits of using LLMs to automate the reasoning tasks related to applying OpenACC directives to existing programs. There are two such reasoning tasks: first, there is the need to identify data-parallel loops among all the loops in programs; the second reasoning task relates to devising the correct directives and clauses to use in the pragma directives for parallel operations and data movement strategies for these data-parallel loops. The two reasoning tasks are considerably different and require different training data. The ACCeLLiuM dataset is designed to train LLMs to address the \emph{second} reasoning task, i.e. devise OpenACC pragmas for loops that are already identified as potentially benefiting from GPU acceleration. With the results from our study, we answer the following research questions:\\

\noindent
\textbf{[RQ1]}: How effective are out-of-the-box LLMs in generating syntactically and semantically correct OpenACC pragmas for data-parallel loops?\\
\textbf{[RQ2]}: To what extent does supervised fine-tuning improve the LLMs' performance?\\
\textbf{[RQ3]}: Is a base model pre-trained on code data (CodeLlama) a better starting point than a general-purpose one (Llama 3.1) for generating OpenACC pragmas?\\
\textbf{[RQ4]}: What are the common error patterns in the pragmas generated by ACCeLLiuM-tuned models?\\

Our experiments reveal a stark contrast between base and fine-tuned LLMs. While base LLMs failed to consistently generate valid OpenACC pragmas when prompted, our ACCeLLiuM-fine-tuned models performed considerably well. The fine-tuned CodeLlama model achieves 50.4\% exact-match accuracy and generates the correct primary directive in the pragma 87.3\% of the time. This improvement provides strong evidence that SFT with a high-quality dataset is a highly effective strategy to train LLMs to effectively learn this complex task. The contributions of this work are as follows: \\
\textbullet A publicly available dataset of OpenACC pragma-loop pairs gathered from public code repositories and a pipeline for supervised fine-tuning.\\
\textbullet An open-source pipeline for curating and structuring such code datasets, enabling further research in this area.\\
\textbullet A family of open-weights, fine-tuned LLMs capable of generating OpenACC pragmas for data-parallel loops.\\
\textbullet An experimental evaluation demonstrating the feasibility and effectiveness of using SFT with a high-quality domain-specific dataset to teach LLMs this relatively difficult task of identifying parallelization opportunities in previously unseen \texttt{for} loops and generating the most appropriate OpenACC pragma for them.\\

The remainder of this paper is organized as follows. Section~\ref{sec:related_work} reviews related work on automated parallelization. Section~\ref{sec:accellium_resource_bundle} details our data curation methodology and the fine-tuning of the ACCeLLiuM models. Section~\ref{sec:experimental_setup} outlines our experimental setup and Section~\ref{sec:results_and_analysis} presents our results that answer our research questions. Finally, we conclude in Section~\ref{sec:conclusion_and_future_work}.

\section{Related Work}
\label{sec:related_work}
Automatically parallelizing sequential programs for GPU execution has been a long-standing challenge and an area of active research in software engineering and HPC. Research efforts in the generation of effective, expert-level OpenACC pragmas have been mostly limited to compiler-centric tools. 

Before LLMs, the generation of OpenACC pragmas depended on static analysis and compiler utilities. Tools like DawnCC~\cite{dawncc_2017}, 
pioneered this space by leveraging static and range analysis to identify safe insertion points for OpenMP and OpenACC pragmas in the program. Although effective on static and structured benchmarks like PolyBench, it is limited by its ability to optimize interprocedural data transfers, often leading to performance bottlenecks in CPU-GPU communication in real-world code. Furthermore, it has difficulty selecting optimal pragmas, leading to poorer and slower performance on the GPU compared to the original sequential code~\cite{WangFarui2021Atod}. Similarly, KernelGen~\cite{Mikushin2014KernelGenT} proposed a compiler/runtime hybrid prototype that automates the migration of legacy numerical models via runtime loop analysis and just-in-time (JIT) compilation. It optimizes GPU utilization by initially allocating most of the program to the GPU, while the CPU manages kernel execution. KernelGen creates a GPU-specific LLVM intermediate representation prior to executing GPU kernels. However, it cannot fully exploit all devices within heterogeneous systems~\cite{auto_migrate_seq_to_hetero}. Its dependence on a specialized and potentially unmaintained compiler infrastructure presents a significant barrier to maintained and practical adoption by developers. Other non-learning-based approaches include grammar-based genetic programming techniques to automatically generate and insert OpenACC directives explored by Bruce and Petke~\cite{Bruce2018RN1}. While demonstrating the potential of evolutionary methods, these methods achieved only modest performance gains (2.44–2.60\% reduction in execution time), falling significantly short of the speed-ups achievable with handwritten OpenACC pragma directives (up to 65.68\%). A common thread across these compiler-centric solutions is their reliance on complex toolchains, which would require developers to acquire a deeper understanding of their internal working in order to set them up in their workflow and use them to obtain desired parallelization results. Furthermore, such tools rely on rigid heuristics for their core functionality, which is why they struggle with the ambiguity and diversity of real-world code, especially in larger production-level codebases.

More recently, the success of Large Language Models (LLMs) in code understanding has inspired a new wave of research focused on generating parallelization directives. This work, however, has been overwhelmingly concentrated on OpenMP for multi-core CPUs. For instance, AutoParLLM~\cite{mahmud_autoparllm_2025} used a Graph Neural Network 
trained on the PerfoGraph representation of source C files from the OMP\_Serial ~\cite{chen_ompserial_dataset_2023} dataset to learn the flow-aware characteristics of programs, such as control flow, data flow, call flow, to guide an LLM to generate parallel code by constructing a GNN-guided OpenMP prompt. This pre-trained GNN is used to discover parallelism opportunities and predict the parallelization patterns of specific OpenMP clauses such as \texttt{private}, \texttt{reduction}, \texttt{do-all}, or \texttt{private} and \texttt{reduction} together. Similarly, Nichols et al.~\cite{hpc_coder_nichols_2024} demonstrated the effectiveness of fine-tuning LLMs like GPT-2, GPT-Neo and PolyCoder on HPC-specific scientific code tasks: parallel code completion, predicting OpenMP pragmas, and relative performance prediction. For OpenMP pragma generation, they fine-tune these models on a smaller dataset of \texttt{for} loops with their OpenMP pragmas from an HPC code dataset~\cite{hpc_coder_nichols_2024}. They evaluated the syntactic correctness of their generated pragmas through textual matching. To check for functional correctness, they perform a string-based matching comparison of the generated pragma and the actual pragma while ignoring the differences that do not contribute to functionality. Their work established the viability of fine-tuning LLMs on HPC-specific code for this domain-specific task of OpenMP pragma generation. This line of research of using smaller, highly specialized models HPC tasks is carried forward by Monocoder~\cite{monocoder_kadosh}, a 0.9B parameter model pre-trained from scratch on the C/C++ HPCorpus dataset. Their evaluations suggest that it outperforms larger models in OpenMP pragma generation, in part by better understanding code structure. OMPGPT~\cite{chen_ompgpt_2024} introduced another compact, domain-specific model tailored for OpenMP. They both demonstrate that compact models trained specifically on HPC code can outperform large general-purpose models like GPT-3.5 on OpenMP pragma generation task.

The sophistication and focus of these research works indicate that there are significant steps in automating OpenMP parallelization by developing highly specialized LLMs. However, there is a gap in research work and resources for automatic OpenACC parallelization using LLMs. The challenge of generating OpenACC pragmas is distinct and, in many ways, more complex, requiring nuanced reasoning about data movement across the heterogeneous memory spaces of CPU-GPU systems. To the best of our knowledge, no prior work has systematically curated a high-quality dataset for this task or developed specialized models to address it. ACCeLLiuM is the first to bridge this gap, providing the foundational dataset and fine-tuned models necessary to bring the power of modern LLMs to OpenACC pragma generation.

\section{The ACCeLLiuM Resource Bundle}
\label{sec:accellium_resource_bundle}
At its core, \textbf{ACCeLLiuM} (Open\textbf{ACC} + \textbf{LLM}) is an end-to-end resource designed to facilitate training LLMs to annotate data-parallel loops with OpenACC programs in C/C++ programs. This resources comprises of (1) a novel dataset of $\approx$ 4,000 OpenACC pragma-loop pairs mined from public GitHub C and C++ repositories curated for Supervised Fine-Tuning (SFT), (2) a family of locally deployable open-weights LLMs built on top of Llama 3.1 70B and CodeLlama 34B fine-tuned using supervised fine-tuning on this dataset, and (3) an open-source pipeline for dataset creation, model fine-tuning, and evaluation.

In this section, we describe our reasoning behind selecting the base models and the process of creating the SFT dataset.

\subsection{\textbf{Dataset Creation}}
\label{sec:dataset_creation_subsection}
To train LLMs to generate correct OpenACC pragmas, our primary objective was to construct a dataset of pragma-loop pairs that accurately represent valid parallelization patterns found in real-world C/C++ code. The curation process involved a multistage pipeline, as summarized in Table~\ref{tab:data_filtering}, designed to systematically source, extract, filter, and format the dataset for supervised fine-tuning. The pipeline consists of the following steps:
\begin{table}
  \centering
  \begin{tabular}{@{} l r @{}}
    \toprule
    \textbf{Processing Step} & \textbf{\# of pragma-loop pairs} \\
    \midrule
    \emph{Initially} & 25,656 \\
    Remove Pairs with Invalid Loops  & 10,503 \\
    Deduplicate                      & 4,033 \\
    \midrule
    \textbf{Total Pragma-Loop Pairs} & \textbf{4,033} \\
    Training Dataset (80\%)          & 3,223 \\
    Testing Dataset (20\%)           & 810 \\
    \bottomrule
  \end{tabular}
  \caption{Dataset Creation through the Steps}
  \label{tab:data_filtering}
\end{table}

\textbf{1. Dataset Sourcing and Pragma-Loop Extraction:} The process starts by using the GitHub Code Search API to search for C/C++ files with "\texttt{\#pragma acc loop}" or "\texttt{\#pragma acc parallel loop}". This initial search yielded 1,509 files containing 30,749 pragma instances. OpenACC pragmas serve two main purposes: data management and loop parallelization. This study focuses on the latter, so from these files we extracted pragma-loop pairs where a pragma directly precedes a \texttt{for} loop. To extract the pragma-loop pairs from the program files, we parsed each source file into an Abstract Syntax Tree (AST) using the \texttt{tree-sitter} tool~\cite{tree-sitter-documentation, baxter_clone_1998}. We then traversed each AST using the \texttt{tree-sitter} tool and regex patterns, locating nodes corresponding to OpenACC pragmas and their immediately succeeding \texttt{for} loop by scanning the pragma node's sibling nodes for \texttt{for\_statement} nodes. To ensure consistent representation for model training, each extracted pragma was normalized by removing redundant spaces and standardizing whitespace within parentheses. This step produced an initial set of 25,656 raw pragma-loop pairs. \\ \\
\textbf{2. Data Filtering and Cleaning:} A manual inspection revealed that the raw dataset contained significant noise. Similarly to the observations of other researchers for OpenMP~\cite{harel_pragformer_2024}, many pragma-loop pairs originated from unit tests in the GCC or LLVM OpenACC compiler suites. Such examples are created solely for testing compiler compatibility and include empty or infinite loops, intentionally erroneous loop bodies, pragmas that did not match the loop structure, and control flow statements incompatible with OpenACC parallelization best practices~\cite{openacc-best-practices-guide}. These examples are unsuitable for training models to generate the correct pragma for real-world code. We therefore implemented a series of filtering steps to enhance the quality of the dataset. We filter out:
\begin{itemize}
    \item Empty or infinite loops (e.g., \verb|for(...)|, \verb|for(...){}|, or \verb|for(...){;}|) as these are not targets for parallelization and thus not suitable training samples.
    \item Loops with incompatible control flow (e.g., \\
    \texttt{break\_statement}, \texttt{goto\_statement}, \\
    \texttt{continue\_statement}, and \texttt{return\_statement}) as they can introduce loop-carried dependencies, force non-sequential execution, cause data races, or imply premature exits. All of these violate the rules of simple data-parallelism and can lead to undefined behavior on the GPU.
\end{itemize}
This filtering process reduced the dataset to 10,503 valid pragma-loop pairs.\\
\textbf{3. Final Dataset Construction and Formatting:} The filtered pairs were de-duplicated based on an exact match comparison of their loop body, resulting in 4,033 unique examples. Each unique pragma was assigned a complexity score based on the number of directives and clauses in it, ensuring that our dataset represents a diverse and balanced range of pragma complexities. The final dataset was shuffled based on this complexity score and evenly distributed into the training set (3,223 pairs, 80\%) and the testing set (810, 20\%). Table~\ref{tab:data_filtering} shows the progression of the size of the dataset through the filtration steps. Table~\ref{tab:pragma-complexity-frequencies} shows the distribution of the pragmas in the final dataset by complexity, while Table~\ref{tab:directive-distribution} shows the distribution of the directive types in the pragmas in the final dataset. Each pragma-loop pair was formatted into a JSONL chat template, as shown in Listing~\ref{lst:single_dataset_entry_exmaple}, as prescribed by the LLM model card~\cite{openai_chat_template}, to support fine-tuning. The format has distinct roles for the system prompt, the user-provided loop, and the assistant's ground-truth pragma response. The test set remains unseen during training.

\begin{table}
  \centering
  \footnotesize
  \begin{tabular}{@{} p{0.55\columnwidth} r @{}}
    \toprule
    \textbf{Pragma Complexity Type}\\[-0.5ex] 
    \textbf{(by Complexity Score)} 
      & \textbf{Frequency} \\
    \midrule
    simple (0--2)       & 1727 \\
    medium (3--5)       & 2290 \\
    complex (6--10)     &   13 \\
    very complex (11+)  &    3 \\
    \bottomrule
  \end{tabular}
  \caption{Pragma Complexity Distribution}
  \label{tab:pragma-complexity-frequencies}
\end{table}

\begin{table}
    \centering
    \begin{tabular}{@{} l r @{} }
        \toprule
        \textbf{Directive Type} & \textbf{Frequency} \\
        \midrule
        loop     & 2565 \\
        parallel & 1262 \\
        kernels  &  137 \\
        serial   &   22 \\
        unknown  &   19 \\
        enter    &   16 \\
        data     &   10 \\
        exit     &    1 \\
        wait     &    1 \\
        \bottomrule
    \end{tabular}
\caption{Directive Type Distribution}
\label{tab:directive-distribution}
\end{table}

\begin{lstlisting}[
  language=json,
  caption={Example of a single JSONL item in the dataset},
  label={lst:single_dataset_entry_example},
  frame=single,
  rulecolor=\color{gray!50},
  backgroundcolor=\color{gray!3},
  numbers=left,
  numberstyle=\tiny\color{gray!70},
  numbersep=6pt,
  xleftmargin=1.2em,
  xrightmargin=1.2em,
  basicstyle=\ttfamily\small,
  showstringspaces=false,
  breaklines=true,
  columns=fullflexible,
  postbreak=\mbox{\textcolor{black}{$\hookrightarrow$}\space}
]
{ "messages": [{
"role": "system",
"content": "You are an expert in crafting optimal OpenACC pragma directives for GPU parallelization.
Your ONLY task is to provide EXACTLY ONE LINE containing ONLY the most appropriate pragma directive, beginning with `#pragma acc`, to replace the placeholder marked by `<TARGET_PRAGMA_LOCATION>` in the given code snippet. Think step by step. Before generating the directive, you must perform a careful and thorough analysis, explicitly considering:
    1. The precise loop structure and all data dependencies to ensure correctness.
    2. Memory access patterns, identifying potential optimizations or hazards.
    3. The most suitable OpenACC directive (`parallel`, `kernels`, `loop`) aligned with identified loop characteristics.
    4. The inclusion of specific clauses (`gang`, `worker`, `vector`, `collapse`, `reduction`, etc.) explicitly chosen to optimize performance, address dependencies, and ensure efficient memory utilization.
Your final output must strictly adhere to these requirements:
- A SINGLE LINE pragma directive starting exactly with `#pragma acc`.
- NO additional text, explanation, comments, or code."}, {
"role": "user",
"content": "
<TARGET_PRAGMA_LOCATION>
for(size_t i=0; i<size; ++i){
    for(size_t j=0; j<size; ++j){
        sum += mat[i*size+j];
        }
    }"}, {
"role": "assistant",
"content":
"#pragma acc parallel loop present(mat[0: size*size]) reduction(+:sum)" }
]}
\end{lstlisting}

\subsection{\textbf{Model Selection and Supervised Fine-Tuning}}
\label{sec:model_selection_and_sft_subsection}
To investigate the comparison between a base model and a code-trained model in this specialized task (\textbf{RQ2}), we selected two distinct open-weights models: (1) Llama 3.1 70B ~\cite{grattafiori2024llama3herdmodels, chen_evaluating_2021, touvron2023llama, zhang2023llamaadapter, gao2023llamaadapterv2}, a general-purpose foundation model, and (2) CodeLlama 34B~\cite{codellama_rozier_2023}, a variant of the Llama 2 model pre-trained on a large corpus of code data. We chose these models because they are freely available with open source licenses and because of their smaller size compared to commercially used models (GPT).

\section{Experimental Setup}
\label{sec:experimental_setup}
\subsection{Inference}
\label{sec:inference_subsection}
To generate a pragma for a given data-parallel loop, the LLMs are provided with a system prompt establishing their role as an OpenACC expert and the code context containing a \texttt{<TARGET\_PRAGMA\_LOCATION>} marker followed by the data-parallel \texttt{for} loop to parallelize. The LLMs' task is to generate the appropriate OpenACC pragma tailored to the data-parallel loop in a single line of text to replace this marker. The prompt, along with a specific example of a loop, can be seen in Listing~\ref{lst:input_prompt_example}. The generated pragmas are then normalized to standardize their whitespace and formatting to ensure a canonical representation for consistent evaluation by comparison with the reference ground-truth pragmas.

\begin{lstlisting}[
  style=mintedlike,
  language=,
  caption={Example of prompt used to generate OpenACC pragmas. The complete system instruction is the same as in Listing~\ref{lst:single_dataset_entry_exmaple}.},
  label={lst:input_prompt_example}
]
"System Prompt": You are an expert in crafting optimal OpenACC pragma directives for GPU parallelization . . .
"Input":
<TARGET_PRAGMA_LOCATION>
for (int j = 0; j < gs0; ++j) {
  sum += val[j];
}
\end{lstlisting}


\subsection{Model Configuration, Fine-tuning, and Hardware}
\label{sec:model_configuration_ft_hardware_subsection}
We performed Supervised Fine-Tuning on Llama 3.1 70B~\cite{grattafiori2024llama3herdmodels} and CodeLlama 34B~\cite{codellama_rozier_2023} models using our OpenACC dataset, consisting of 3,223 examples similar to that in Listing~\ref{lst:input_prompt_example}. Typically, larger models often yield superior performance but require significantly more computational resources such as GPU VRAM memory or training time. Therefore, we used relatively smaller models that could fit on a single GPU. We used Unsloth's Supervised Fine-Tuning Trainer (\texttt{SFTTrainer}) with parameter efficient fine-tuning using the QLoRA~\cite{peft, dettmers_qlora_2023} method since this approach substantially reduces memory requirements by quantizing the base model and only training a small set of adapter weights. We used the Unsloth library for this task, as it provides an optimized implementation of QLoRA to efficiently fine-tune and evaluate both models on our test dataset using resource-constrained hardware~\cite{unsloth}. Training was conducted on a $1\times$ NVIDIA H100 $80GB$ GPU for 3 epochs, at \texttt{bf16} precision. We used AdamW optimizer with a learning rate of $6.0e-5$, a warm-up ratio of $0.1$, and a cosine annealing schedule.

\subsection{Evaluation Metrics}
\label{sec:evaluation_metrics_subsection}
To answer our research questions, we empirically evaluate the ACCeLLiuM models by comparing the quality of the pragma generated for each data-parallel loop with its original human-written reference pragma. Assessing the quality of the LLM-generated pragmas requires a nuanced definition of "correctness." A generated pragma must be both syntactically valid for the compiler and semantically appropriate to parallelize the operations. We therefore designed a two-pronged evaluation protocol: (1) correct use of individual pragma components, namely, the primary directive and the clauses~\cite{chen_evaluating_2021} and (2) syntactic validity.

We assess the model-generated pragmas' semantic alignment with the human-written ground-truth reference pragmas using a suite of metrics. \textbf{Exact match accuracy} measures the model's ability to correctly reproduce the pragma exactly as the reference pragma. In contrast, the \textbf{Levenshtein similarity}, which is based on the Levenshtein edit distance~\cite{wiki_levenshtein_distance} between the characters of the two pragma strings, credits partial similarities between the generated and reference pragmas. This allows for a less strict assessment than an exact match.

We normalize the Levenshtein edit distance to a range between 0 and 1, where 1 is identical. Scores greater than 0.8 indicate that the generated pragmas are structurally very similar to the reference, with correct directive, clauses, and clause parameters, and only minor variations. Scores between 0.5 and 0.8 reflect moderate similarity, suggesting some errors in clauses or clause parameters. Scores below 0.5 indicate that the generated pragma is largely different from the reference.

These two metrics compare entire strings and therefore fail to capture clause reorderings that alter the string despite semantic equivalence. For example, the OpenACC pragmas in Listing~\ref{lst:reordered_clauses} provide the same semantic instructions to the compiler on how to parallelize the associated data-parallel loop, despite having clauses in a different order. In order to better capture these situations, we used two additional metrics to evaluate specific components. \textbf{Directive-type match} evaluates whether the model selected the correct primary directive in the pragma, such as \texttt{loop}, \texttt{parallel}, \texttt{kernels}, or \texttt{data}. We report macro-averaged precision, recall, and F1 score for all unique directive types~\cite{monocoder_kadosh}, as can be seen in Table~\ref{tab:p_f1_r}. \textbf{Clause-wise Jaccard similarity} is another component-level evaluation metric that measures the overlap in the set of clauses between the generated and reference pragmas. It ignores the order of the clauses to capture functional correctness, as defined by the OpenACC specification~\cite{openacc_specification_2021}, even with syntactic variations, where string-based metrics fail. Scores above 0.8 indicate high clause accuracy; 0.5–0.8 suggests partial yet acceptable correctness; below 0.5 indicates poor clause selection. The model may also add valid clauses that are not in the reference. For example, for the reference and generated pragmas in Listing~\ref{lst:reordered_and_added_clause}, the Jaccard score would be $2/3 = 0.667$ since two thirds of the clauses overlap between the two pragmas. 

\begin{lstlisting}[
  style=mintedlike,
  language=C,
  caption={Semantically identical pragmas with reordered clauses},
  label={lst:reordered_clauses}
]
#pragma acc parallel loop present(val[0:gs0]) reduction(+ : sum)

#pragma acc parallel loop reduction(+ : sum) present(val[0:gs0])
\end{lstlisting}

\begin{lstlisting}[
  style=mintedlike,
  language=C,
  caption={Pragmas with reordered clauses, and an additional clause, both semantically valid with the second one offering potentially better parallelization},
  label={lst:reordered_and_added_clause}
]
#pragma acc parallel loop copyin(a) present(b)

#pragma acc parallel loop present(b) copyin(a) reduction(+:c)
\end{lstlisting}

\begin{lstlisting}[
  style=mintedlike,
  language=C,
  numbers=left,
  caption={Minimal Compilable Unit (MCU) Example},
  label={lst:mcu_example}
]
// Minimal C MCU for OpenACC pragma validation
// #include <openacc.h> // May not be strictly needed for syntax check with -c
void test_loop_pragma_mcu() {
    int n = 100;
    double a, b, c; // Variables used in the loop
    double sum_val = 0.0;          // Variable for reduction
    int i;                         // Loop index, often implicitly private
    // Reference or Generated Pragma
    #pragma acc parallel loop reduction(+:sum_val) private(i) copyin(a[0:n],b[0:n]) copyout(c[0:n])
    for (i = 0; i < n; ++i) {
        c[i] = a[i] + b[i];
        sum_val += c[i];
    }
}
/*
// To compile for syntactic/semantic check, with no pragma (creates minimal_c_mcu.o if successful):
// nvc -c minimal_c_mcu.c -o minimal_c_mcu.o
// To compile for syntax/semantic check, with pragma (creates minimal_c_mcu.o if successful):
// nvc -c -acc -Minfo=accel minimal_c_mcu.c -o minimal_c_mcu.o
*/
\end{lstlisting}

To evaluate the syntactic validity of LLM-generated pragmas for an OpenACC-compliant compiler (such as \texttt{nvc}), we generated minimal compilable units (MCU) for every pragma-loop pair in the test dataset using the Gemini 2.5 Pro Model~\cite{gemini_pro}. The prompt used to generate these MCUs is provided as supplementary material to respect the space constraints of the paper. An MCU is the smallest possible self-contained program that includes the essential \texttt{\#include} directives, the target data-parallel loop, and minimal but sufficient variable declarations for all variables referenced within the loop body and in any pragma clauses, including dummy initializations that might be necessary for the compiler to perform semantic checks. It provides just enough context for the compiler's front-end (parser and semantic analyzer) to create a compilable object of the MCU to verify its syntactic correctness. This avoids compilation failures due to missing definitions, unresolved external symbols, or complex dependencies present in the larger application, allowing the focus to remain squarely on the pragma's validity. A successful compilation serves as a definitive and objective measure of syntactic correctness. We used the Gemini 2.5 Pro model for this task because of its ability to infer variable initializations and context, which is challenging for a purely static analysis script. Listing~\ref{lst:mcu_example} is an example of an MCU. Each MCU is compiled in three combinations: 
\begin{enumerate}
    \item \textbf{MCU with no pragma}: Compiled for serial execution, without activating the OpenACC standard in the compiler terminal command. Creates a baseline of whether the \texttt{for} loop is compilable or not. Subsequent MCU combinations that include reference or generated pragmas are only compiled if this first MCU combination compiles successfully.
    
    \item \textbf{MCU with reference pragma}: Compiled for parallel execution with the OpenACC standard activated (\texttt{-acc}) in the compiler command to instruct the compiler to apply the OpenACC syntax rules during compilation.
    
    \item \textbf{MCU with generated pragma}: Compiled for parallel execution with the Openacc standard activated \texttt{-acc} in the compiler command. This is the final compilability check for the generated pragmas. 
\end{enumerate}

High Jaccard with low Levenshtein score means correct clause names but formatting or argument variations. Low Jaccard but high syntax validity score suggests valid but contextually wrong clauses. Jaccard thus isolates the quality of the selection of clauses, showing whether the model captures key parallelism strategies for the given data-parallel loop and uses the correct clauses effectively. We also measure the extraction failure rate to track when the model outputs unusable text.

\section{Results and Analysis}
\label{sec:results_and_analysis}
\begin{figure}[t]
  \centering
  \includegraphics[width=\columnwidth]{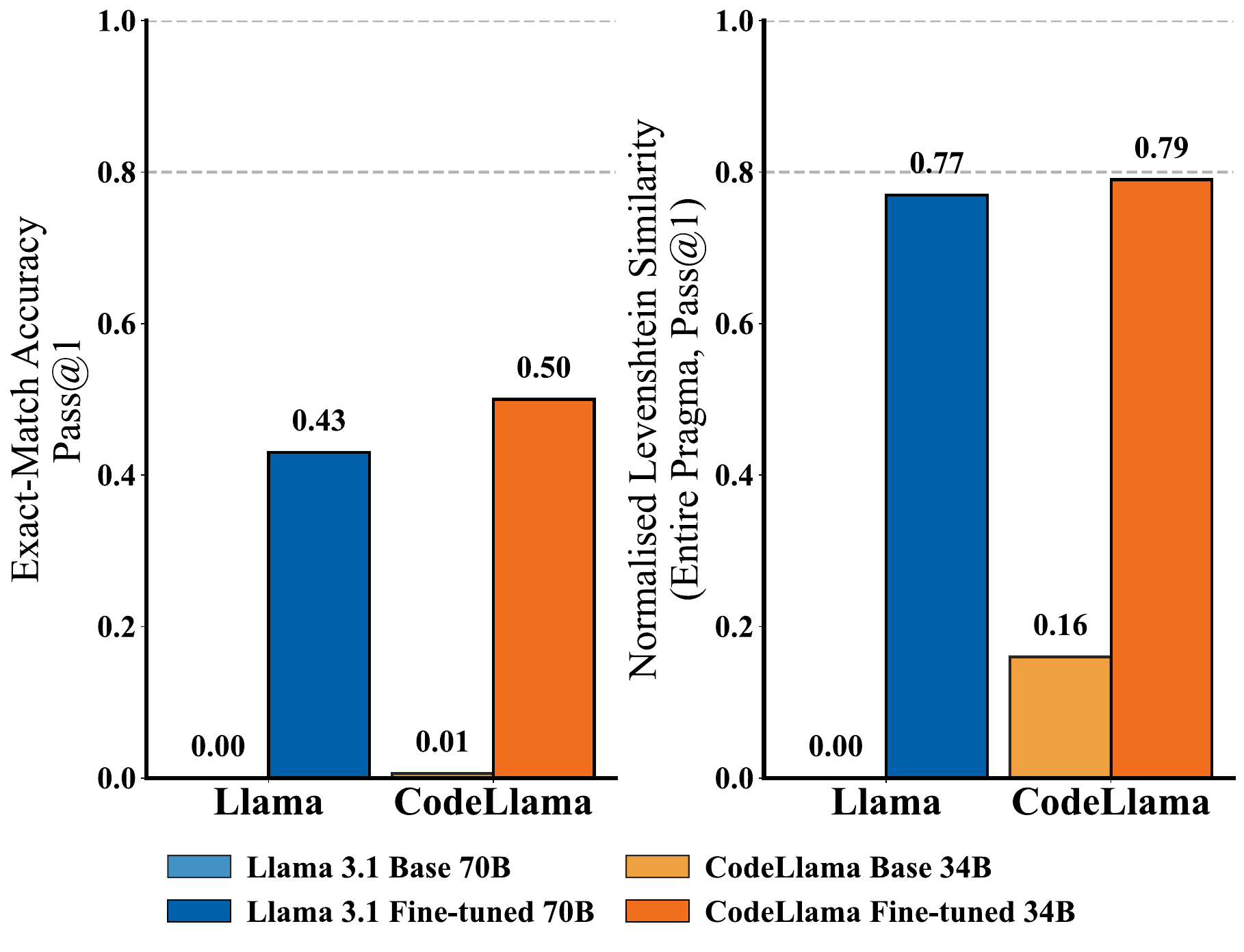}
  \caption{(Left) Comparison of the models in Exact Match Accuracy (\%). (Right) Comparison of the models in Levenshtein Similarity.}
  \label{fig:exact_match_and_levenshtein_similarity}
\end{figure}

\begin{figure}[t]
  \centering
  \includegraphics[width=1.0\columnwidth]{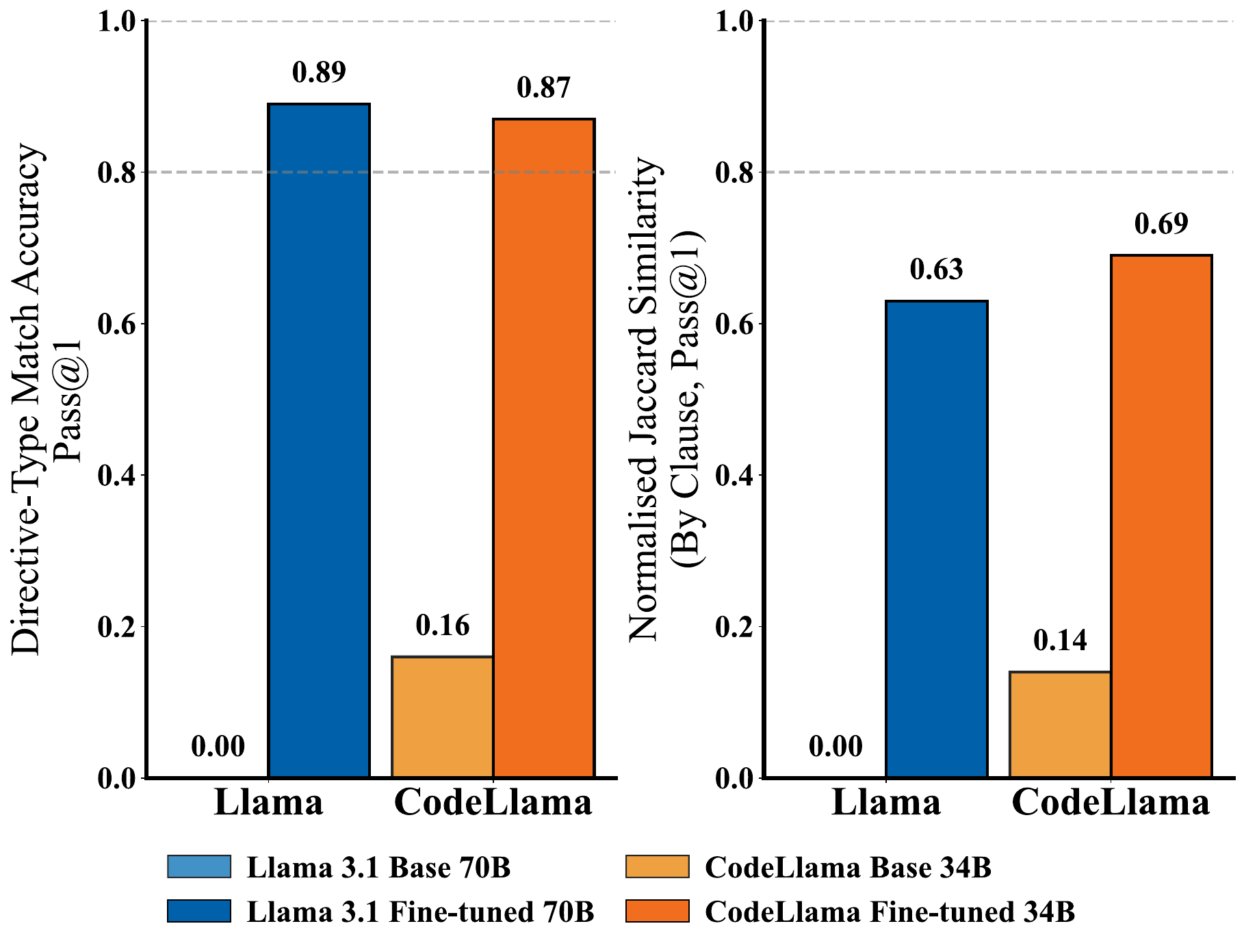}
  \caption{(Left) Comparison of the models in Directive-Type Match Accuracy (\%). (Right) Comparison of the models in clause-wise Jaccard similarity.}
  \label{fig:directive_type_match_and_jaccard_similarity}
\end{figure}

The charts in Figures~\ref{fig:exact_match_and_levenshtein_similarity}, \ref{fig:directive_type_match_and_jaccard_similarity}, and Table~\ref{tab:p_f1_r} summarize our findings. This section explains these results in light of the four research questions outlined in Section~\ref{sec:introduction}\footnote{All values reported are Pass@1}.

\begin{table}
  \centering
  \begin{tabular}{@{} l r r r @{}}
    \toprule
    \textbf{Model} & \textbf{P (\%)} & \textbf{R (\%)} & \textbf{F$_1$ (\%)} \\
    \midrule
    Llama 3.1--Base & 0.00 & 0.00 & 0.00 \\
    Llama 3.1--ACCeLLiuM Fine‑tuned & 55.24 & 48.80 & 49.45 \\
    CodeLlama--Base & 4.80 & 3.70 & 4.20 \\
    CodeLlama--ACCeLLiuM Fine‑tuned & 45.20 & 33.80 & 37.47 \\
    \bottomrule
  \end{tabular}
  \caption{Precision, Recall, and F$_1$‑Score for Directive Type Prediction}
  \label{tab:p_f1_r}
\end{table}

\subsection{RQ1 Results}
\label{sec:rq1_subsection}
\textbf{RQ1: How effective are out-of-the-box LLMs in generating syntactically and semantically correct OpenACC pragmas for data-parallel loops?} To address \textbf{RQ1}, we evaluate how well the base Llama 3.1 70B and CodeLlama 34B models generate OpenACC pragmas for each data parallel loop in the test dataset. We score the pragmas they generate with the suite of metrics described in Section~\ref{sec:evaluation_metrics_subsection}. For every data-parallel loop in the test dataset, we provide the models with the prompt in the format seen in Listing~\ref{lst:input_prompt_example} and ask them to generate a single OpenACC pragma line. 

For semantic correctness, the base models performed poorly. As seen in Figure~\ref{fig:exact_match_and_levenshtein_similarity}, left side, base Llama 3.1 failed to generate any correct pragmas (0\% Exact Match Accuracy), while the base CodeLlama achieved an exact match accuracy of only 0.01 out of 1.0. CodeLlama-generated pragmas reached an average Levenshtein similarity of just 0.16 with the reference pragmas (Figure~\ref{fig:exact_match_and_levenshtein_similarity}, right side).

For directive-type match, pragmas generated by base CodeLlama model only scored a normalized score of 0.16 out of 1.0 (Figure~\ref{fig:directive_type_match_and_jaccard_similarity}, left side). CodeLlama achieved a low mean Jaccard similarity score of 0.14, indicating minimal overlap with the clauses in the ground-truth reference pragma (Figure~\ref{fig:directive_type_match_and_jaccard_similarity}, right side). 

These results collectively show that base models struggle to produce correct OpenACC pragmas.

\subsection{RQ2 Results}
\label{sec:rq2_subsection}
\textbf{RQ2: To what extent does supervised fine-tuning improve the LLMs' performance?} After undergoing supervised fine-tuning on our dataset, we evaluated the fine-tuned models using the test dataset, the same way as we did to evaluate their base versions to answer RQ1. For exact match accuracy, the fine-tuned Llama 3.1 achieved 43\% accuracy, and the fine-tuned CodeLlama achieved 50\% accuracy (Figure~\ref{fig:exact_match_and_levenshtein_similarity}, left side). Furthermore, pragmas generated by our fine-tuned models also reached a high average Levenshtein similarity score of 0.77 (Llama 3.1) and 0.79 (CodeLlama) (Figure~\ref{fig:exact_match_and_levenshtein_similarity}, right side), indicating that many non-exact matches are structurally very similar to the reference pragma. 

For directive-type match, our ACCeLLiuM-fine-tuned\\
Llama 3.1 and CodeLlama models achieved 89\% and 87\% accuracy, respectively. Despite being older and smaller, the fine-tuned CodeLlama performs nearly as well as the fine-tuned Llama 3.1 (Figure~\ref{fig:directive_type_match_and_jaccard_similarity}, left side). Our ACCeLLiuM-fine-tuned models achieve a high average clause-wise Jaccard similarity of 0.63 (Llama 3.1) and 0.69 (CodeLlama), as seen in Figure~\ref{fig:directive_type_match_and_jaccard_similarity} (right side).

We measure the \textbf{syntactic validity} of the data parallel loops, the reference pragmas, and the generated pragmas using minimal compilable units (MCU) as detailed in Section~\ref{sec:evaluation_metrics_subsection}. This process first filters for MCUs with compilable \texttt{for} loops with no pragma present, resulting in 762 valid test cases from the original 810. This initial compilability check of MCUs with only the data-parallel loops, without activating the OpenACC framework during compilation, from the test dataset is essential to determine how many of them were actually syntactically correct. Our intuition that the Gemini model may produce some syntactically invalid MCUs was correct. This forms a baseline of 762 compilable loops from our test dataset. For each of these 762 compilable MCUs, two new MCUs are created--one with the reference pragma and another with the generated pragma replacing the \texttt{<TARGET\_PRAGMA\_LOCATION>} in the MCU. The syntactic validity of the pragmas is determined by whether the MCU with that pragma compiles successfully when compiled with the OpenACC framework (using the \texttt{-acc} tag in the compilation command).

\begin{figure}
  \centering
  \includegraphics[width=0.80\columnwidth]{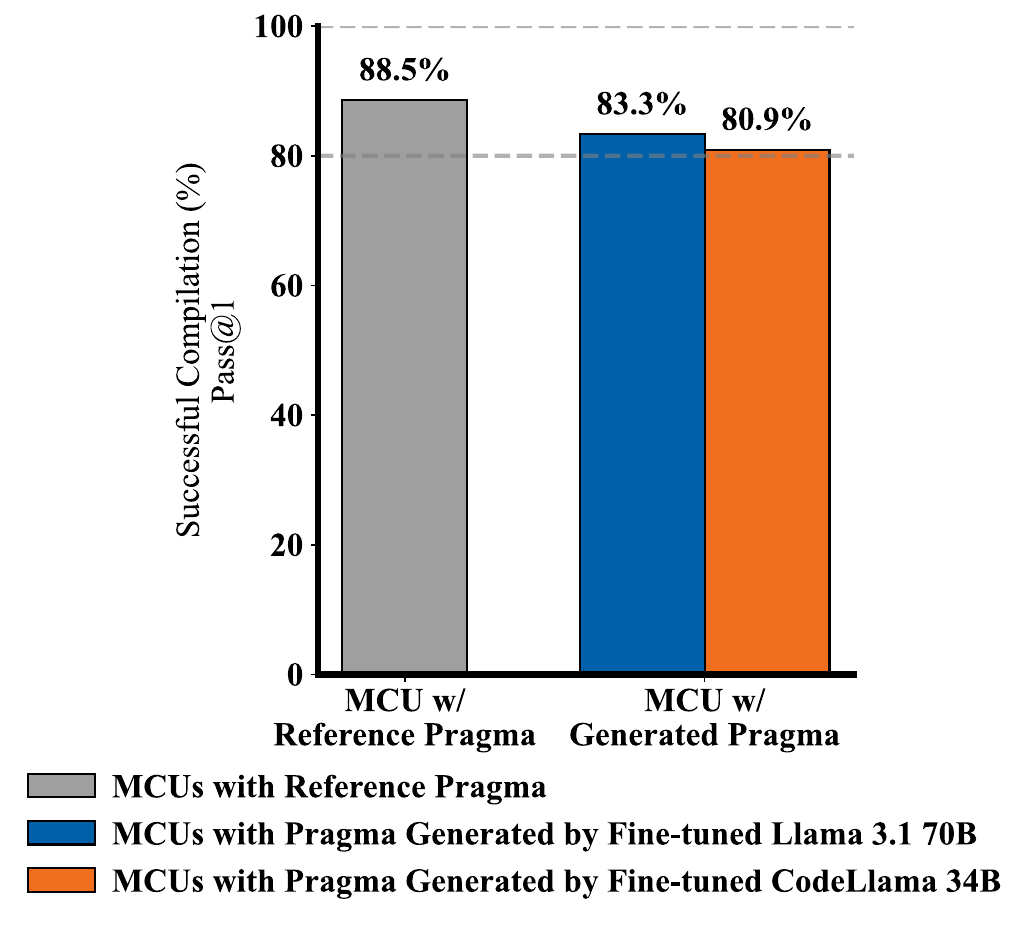}
  \caption{MCU Compilation Success Rate (\%).}
  \label{fig:compilation_eval_grouped}
\end{figure}

Pragmas generated by ACCeLLiuM-fine-tuned models demonstrate high syntactic correctness in these valid test cases. Against this baseline, our ACCeLLiuM-fine-tuned Llama 3.1 and CodeLlama models generated pragmas that compiled successfully in 83.3\% (635 out of 762) and 80.9\% (617 out of 762) of the cases, respectively (Figure~\ref{fig:compilation_eval_grouped}). These rates are comparable to the 88.5\% (674 out of 762) compilation rate of human-written reference pragmas, suggesting a high degree of syntactic reliability. This indicates that the models have effectively learned the grammatical structure of the OpenACC pragmas and the use of appropriate directives, clauses, and clause variables tailored to the parallelization patterns in a given data-parallel loop. Notably, even 11.5\% (88 out of 762) of the human-written reference pragmas in our test set failed to compile, suggesting the presence of syntactic errors in the original ground-truth data.

Both our fine-tuned models consistently generated syntactically valid and semantically relevant pragmas for a wide range of complex loops, successfully generating complex, multi-clause pragmas that require a deeper understanding of the loop's data dependencies and operations. As demonstrated by the example in Listing~\ref{lst:correct_pragma_complex_loop}, the models correctly identified the nested loop structure suitable for \texttt{collapse(2)}, and also correctly inferred the data movement semantics: array \texttt{a} is an input (\texttt{copyin}) and array \texttt{b} is the output (\texttt{copyout}). This shows that the models learned the necessary reasoning for loop parallelization.

In summary, our evaluation confirms that supervised fine-tuning of the base models using our ACCeLLiuM dataset dramatically improves the performance of the models.

\begin{lstlisting}[
  style=mintedlike,
  language=,
  caption={Correct Pragma Generation for a Complex Loop by Fine-tuned CodeLlama 34B},
  label={lst:correct_pragma_complex_loop}
]
Input: 
for (j = 1; j < n - 1; j++) {
  for (i = 1; i < n - 1; i++) {
    b[i][j] = 0.2 * (a[i][j] + a[i][j-1] + a[i][j+1] + a[i+1][j] + a[i-1][j]);
  } 
}

Generated Pragma (Exact Match): 
#pragma acc parallel loop collapse(2) copyin(a) copyout(b)
\end{lstlisting}

\subsection{RQ3 Results}
\label{sec:rq3_subsection}
\textbf{RQ3: Is a base model pre-trained on code data (CodeLlama) a better starting point than a general-purpose one (Llama 3.1) for generating OpenACC pragmas?} In the initial evaluation of the base versions of the models, CodeLlama performs the same as Llama 3.1 in the syntactic validity and exact match accuracy metrics, producing no considerable syntactically or semantically correct pragmas. The base CodeLlama model performs better than the base Llama 3.1 model in Levenshtein similarity, directive-type match accuracy and Jaccard similarity, as seen in Figures~\ref{fig:exact_match_and_levenshtein_similarity} (right side) and~\ref{fig:directive_type_match_and_jaccard_similarity}. With supervised fine-tuning, the fine-tuned CodeLlama performs slightly better than the fine-tuned Llama 3.1 in exact match accuracy, Levenshtein similarity, and Jaccard similarity metrics (Figures~\ref{fig:exact_match_and_levenshtein_similarity},~\ref{fig:directive_type_match_and_jaccard_similarity} (right side)). The fine-tuned Llama 3.1 performs slightly better than the fine-tuned CodeLlama in directive-type match accuracy (Figure~\ref{fig:directive_type_match_and_jaccard_similarity} (left side)) and syntactic validity of the generated pragmas (Figure~\ref{fig:compilation_eval_grouped}).

Given these results, the answer to this research question is inconclusive. On the one hand, the code model does not seem to bring significant advantages over the general-purpose model: both models perform comparably, with only slight differences. On the other hand, CodeLlama 34B is an older and smaller model than Llama 3.1 70B, and performs similarly, so the fact that it was trained on code may be a cause of its being at par with a newer and larger general-purpose model.

\subsection{RQ4 Results} 
\label{sec:rq4_subsection}
\textbf{RQ4: What are the common error patterns in the pragmas generated by the ACCeLLiuM-fine-tuned models?} The most significant error category across both supervised fine-tuned models is \textbf{Incorrect or Partially Correct Clause Generation}, accounting for the majority of the failures where the primary directive was correctly identified. The models occasionally get confused with data clauses, such as \texttt{copyin} and \texttt{present}, generating pragmas with correct primary directive but incorrect clauses. For example, for a \texttt{for} loop that uses an array \texttt{a} that is expected to be already in the GPU memory, with reference pragma \texttt{\#pragma acc parallel loop present(x)} in some cases they generated \texttt{\#pragma acc parallel loop copyin(x)}. This is a semantic error as \texttt{copyin(x)} instructs the compiler at runtime to transfer data from the CPU memory to GPU memory, which is unnecessary is the data is already \texttt{present} in the GPU. This can lead to redundant data transfers and performance degradation. This error likely occurs because the models are only given the data-parallel loops as input during fine-tuning, not the code before it, which could serve as additional context. The dataset limits the input to the loop to avoid confusion and maintain consistency. In another example, for an input loop that reads from two arrays(\texttt{b} and \texttt{c}) and writes to one (\texttt{a}) with reference pragma \texttt{\#pragma acc parallel loop present(b,c) copyout(a)}, the models generated \texttt{\#pragma acc parallel loop present(b) copyout(a)}. The models correctly identified the overall parallel structure and the output array \texttt{a}, but missed including the array \texttt{c} in the \texttt{present} clause. This is "near miss" is a non-trivial error that would likelycause a compiler error or runtime failure, suggesting that the models' attention to all variables in a complex expression may be a point of weakness.

Another common error involves \textbf{functionally correct pragmas with reordered clauses}, as discussed earlier in Subsection~\ref{sec:evaluation_metrics_subsection}. These cases highlight the limitations of string-based metrics, such as exact match accuracy, and justify the use of clause-wise Jaccard similarity. Furthermore, despite strong directive-type accuracies (89\% for Llama 3.1 and 87\% for CodeLlama), directive-choice errors still occurred. In some data-parallel loops, with a reference pragma \texttt{\#pragma acc parallel loop}, the models generated \texttt{\#pragma acc kernels}. Although syntactically valid, \texttt{kernels} is less explicit and generally not preferred over \texttt{parallel loop} which is often more explicit and preferred for such loops. This suggests that the models have learned multiple ways to express parallelism, but have not yet perfected the nuances of choosing the optimal one in every context.

We analyzed 405 instances (50\% of 810 test cases) generated by the CodeLlama model and 462 instances (57\% of 810 test cases) generated by the Llama 3.1 model where the pragmas did not exactly match the reference. The complete error distribution is shown in Table~\ref{tab:error_comparison}. We categorize their errors as follows:\\
\textbf{1.\;Directive Choice Errors:} These are cases where the model selected an incorrect primary directive. CodeLlama and Llama 3.1 exhibit directive choice failure in 13\% and 11\% of the test cases, respectively, corresponding to $810 \times \left(1 - 0.87\right) = 810 \times 0.13 = 105.3 \approx \textbf{105}$ and $810 \times \left(1 - 0.89\right) = 810 \times 0.11 = 89.1 \approx \textbf{89}$ errors.\\
\textbf{2. Clause Errors, with Correct Directive:} These are errors where the directive was correct, but the clauses were not an exact match. For CodeLlama, the clause-error count is\\
\(\text{Count(Clause Errors)} = E_{\text{total}} - \text{Count(Directive Error)} = 405 - 105 = \textbf{300}\); and for Llama 3.1 it is \(\text{Count(Clause Errors)} = E_{\text{total}} -
\text{Count(Directive Error)} = 462 - 89 = \textbf{373}\). We further subdivide these clause errors:
    \begin{enumerate}[label=\alph*.]
        \item \textbf{Clause Reordering (Semantically Correct):} Cases where the generated pragmas are functionally identical to the reference but with reordered clauses, such as the example in Listing~\ref{lst:reordered_clauses}. We identify these non-exact matches with a Jaccard similarity of 1.0, thus accounting for 20\% of clause errors by the CodeLlama model and 15\% of clause errors by the Llama 3.1 model. \(\text{Count(Reordering)} = 300 \times 0.20 = 60\) for CodeLlama. \(\text{Count(Reordering)} = 373 \times 0.15 = 56\) for Llama 3.1.
        \item \textbf{Major Clause Errors:} The directive is correct, yet one or more clauses are incorrect, missing, or sub-optimal (Jaccard similarity \(<\) 0.5). This includes errors such as using \texttt{copyin} instead of \texttt{present}. These account for 20\% (60 out of 300) of CodeLlama, and 30\% (112 out of 373) of Llama 3.1 clause errors.
        \item \textbf{Minor Clause Errors:} The directive is correct, and there is significant overlap between the clauses of the reference and generated pragmas (Jaccard similarity $>$ 0.5), but a clause may be partially correct or incomplete. This creates "near miss" situations due missing clause parameters or inclusion of redundant clauses like \texttt{private}. An example is missing adding the variable \texttt{c} in a \texttt{present} clause which was present in the reference pragma in th example mentioned earlier in this section. This largest sub‑category comprises 60\% (180 out of 300) of CodeLlama and 55\% (205 out of 373) of Llama 3.1 clause errors. 
    \end{enumerate}

\begin{table}
\setlength\tabcolsep{0.5pt}    
\small
\begin{tabular}{@{}L{0.38\linewidth} rr@{}} 
\toprule
\textbf{Error Category} & \textbf{Fine‑tuned CodeLlama} & \textbf{Fine‑tuned Llama 3.1} \\ \midrule
\textbf{Total Non‑Exact Matches} & \textbf{405(50\% of 810)} & \textbf{462(57\% of 810)} \\ \midrule
\textbf{Directive Choice Error}  & \textbf{105(13\% of 810)} & \textbf{89(11\% of 810)} \\
\textbf{Clause Error (Correct Directive)} & \textbf{300(37\% of 810)} & \textbf{373(46\% of 810)} \\
\subrowindent a.\;Clause Reordering  &  60(20\% of 300) &  56(15\% of 373) \\
\subrowindent b.\;Major Clause Error &  60(20\% of 300) & 112(30\% of 373) \\
\subrowindent c.\;Minor Clause Error & 180(60\% of 300) & 205(55\% of 373) \\
\bottomrule
\end{tabular}
\caption{Comparison of Error Categories between Fine‑Tuned Models}
\label{tab:error_comparison}
\end{table}

\section{Conclusion and Future Work}
\label{sec:conclusion_and_future_work}
In this paper, we addressed the challenge of automating directive-based GPU parallelization by introducing ACCeLLiuM, an end-to-end resource bundle featuring a novel dataset of approximately 4,000 real-world OpenACC pragma-loop pairs and a family of fine-tuned LLMs. Our work confronts a critical gap in the literature, where the focus has predominantly been on OpenMP pragma generation, leaving the more complex task of OpenACC generation largely unexplored.

Our empirical evaluation demonstrates that general-purpose LLMs are largely incapable of generating correct OpenACC pragmas out-of-the-box. However, after supervised fine-tuning on our curated dataset, the models show improvement. The ACCeLLiuM-fine-tuned models achieve high syntactic validity, with over 80\% of generated pragmas compiling successfully, and strong semantic correctness, identifying the correct primary directive in up to 89\% of cases and achieving an exact-match accuracy of 50\%. Our analysis of error patterns reveals that most remaining inaccuracies are "near misses" related to clause selection or missing clause parameters, often stemming from the lack of full-program context. These results provide strong evidence that domain-specific fine-tuning is a highly effective strategy for teaching LLMs the nuanced reasoning required for this complex HPC task.

The primary implication of our work is that it lowers the barrier to GPU acceleration for developers and scientists who may not be experts in parallel programming. By providing both a foundational dataset and open-weights models, we establish a reproducible benchmark for further research into LLM-driven code parallelization.

We acknowledge the limitations of our current study. Our approach focuses on annotating single, pre-identified data-parallel loops, and our evaluation is static, based on syntactic and semantic correctness rather than dynamic performance measurements. The models' most common errors--incorrect data clauses--are likely attributable to the lack of inter-procedural context.

This leads to several promising avenues for future work. The most critical next step is to perform a dynamic evaluation by integrating our models into a compile-and-run feedback loop with established HPC benchmark suites, such as SPEC ACCEL, to measure real-world speedup and functional correctness. We also plan to extend the models' capabilities to reason about larger code regions, enabling the generation of more complex pragmas for more involved parallelization tasks. Finally, exploring reinforcement learning strategies to better align the pragmas with desired parallelization optimizations using data from compiler diagnostics or static analysis feedback represents a compelling direction for creating more robust and interactive parallelization assistants.


\bibliographystyle{IEEEtran}
\bibliography{bibliography}

\begin{thebibliography}{10}
\providecommand{\url}[1]{#1}
\csname url@samestyle\endcsname
\providecommand{\newblock}{\relax}
\providecommand{\bibinfo}[2]{#2}
\providecommand{\BIBentrySTDinterwordspacing}{\spaceskip=0pt\relax}
\providecommand{\BIBentryALTinterwordstretchfactor}{4}
\providecommand{\BIBentryALTinterwordspacing}{\spaceskip=\fontdimen2\font plus
\BIBentryALTinterwordstretchfactor\fontdimen3\font minus \fontdimen4\font\relax}
\providecommand{\BIBforeignlanguage}[2]{{%
\expandafter\ifx\csname l@#1\endcsname\relax
\typeout{** WARNING: IEEEtran.bst: No hyphenation pattern has been}%
\typeout{** loaded for the language `#1'. Using the pattern for}%
\typeout{** the default language instead.}%
\else
\language=\csname l@#1\endcsname
\fi
#2}}
\providecommand{\BIBdecl}{\relax}
\BIBdecl

\bibitem{nvidia_cuda_programming_guide}
NVIDIA, ``Nvidia cuda c++ programming guide -- docs.nvidia.com,'' \url{https://docs.nvidia.com/cuda/cuda-c-programming-guide/index.html}, [Accessed 11-06-2025].

\bibitem{karimi2010performance}
K.~Karimi, N.~Dickson, and F.~Hamze, ``A performance comparison of cuda and opencl,'' \emph{Computing Research Repository - CORR}, vol. arXiv:1005.2581, 05 2010.

\bibitem{fang_performance_comparison_cuda_opencl}
J.~Fang, A.~L. Varbanescu, and H.~Sips, ``A comprehensive performance comparison of cuda and opencl,'' in \emph{2011 International Conference on Parallel Processing}, 2011, pp. 216--225.

\bibitem{automatic_parallelism_data_dependency}
\BIBentryALTinterwordspacing
E.~Urbach. (2012, Apr.) Automatic parallelism and data dependency. Archived 14Jul2014; accessed 21Jul2025. [Online]. Available: \url{https://web.archive.org/web/20140714111836/http://blitzprog.org/posts/automatic-parallelism-and-data-dependency}
\BIBentrySTDinterwordspacing

\bibitem{programming_massively_parallel_processors_book}
D.~B. Kirk and W.-m.~W. Hwu, \emph{Programming Massively Parallel Processors: A Hands-on Approach}, 1st~ed.\hskip 1em plus 0.5em minus 0.4em\relax San Francisco, CA, USA: Morgan Kaufmann Publishers Inc., 2010.

\bibitem{openacc_specification_2021}
OpenACC, ``openacc.org,'' \url{https://www.openacc.org/sites/default/files/inline-images/Specification/OpenACC-3.2-final.pdf}, [Accessed 10-06-2025].

\bibitem{parallel_programming_with_openacc_book}
R.~Farber, \emph{Parallel Programming with OpenACC}, 1st~ed.\hskip 1em plus 0.5em minus 0.4em\relax San Francisco, CA, USA: Morgan Kaufmann Publishers Inc., 2016.

\bibitem{mahmud_autoparllm_2025}
\BIBentryALTinterwordspacing
Q.~I. Mahmud, A.~TehraniJamsaz, H.~D. Phan, L.~Chen, M.~Capot{\u{a}}, T.~L. Willke, N.~K. Ahmed, and A.~Jannesari, ``{A}uto{P}ar{LLM}: {GNN}-guided context generation for zero-shot code parallelization using {LLM}s,'' in \emph{Proceedings of the 2025 Conference of the Nations of the Americas Chapter of the Association for Computational Linguistics: Human Language Technologies (Volume 1: Long Papers)}, L.~Chiruzzo, A.~Ritter, and L.~Wang, Eds.\hskip 1em plus 0.5em minus 0.4em\relax Albuquerque, New Mexico: Association for Computational Linguistics, Apr. 2025, pp. 11\,821--11\,841. [Online]. Available: \url{https://aclanthology.org/2025.naacl-long.593/}
\BIBentrySTDinterwordspacing

\bibitem{unveiling_parallelism_in_serial_code}
\BIBentryALTinterwordspacing
Z.~Li, R.~Atre, Z.~Huda, A.~Jannesari, and F.~Wolf, ``Unveiling parallelization opportunities in sequential programs,'' \emph{Journal of Systems and Software}, vol. 117, pp. 282--295, 2016. [Online]. Available: \url{https://www.sciencedirect.com/science/article/pii/S016412121630005X}
\BIBentrySTDinterwordspacing

\bibitem{Li2013DiscoveryOP}
Z.~Li, A.~Jannesari, and F.~Wolf, ``Discovery of potential parallelism in sequential programs,'' pp. 1004--1013, 2013.

\bibitem{dawncc_2017}
\BIBentryALTinterwordspacing
G.~Mendon\c{c}a, B.~Guimar\~{a}es, P.~Alves, M.~Pereira, G.~Ara\'{u}jo, and F.~M. Q.~a. Pereira, ``Dawncc: Automatic annotation for data parallelism and offloading,'' \emph{ACM Trans. Archit. Code Optim.}, vol.~14, no.~2, May 2017. [Online]. Available: \url{https://doi.org/10.1145/3084540}
\BIBentrySTDinterwordspacing

\bibitem{Mikushin2014KernelGenT}
\BIBentryALTinterwordspacing
D.~Mikushin, N.~Likhogrud, E.~Z. Zhang, and C.~Bergstrom, ``Kernelgen -- the design and implementation of a next generation compiler platform for accelerating numerical models on gpus,'' \emph{2014 IEEE International Parallel \& Distributed Processing Symposium Workshops}, pp. 1011--1020, 2014. [Online]. Available: \url{https://api.semanticscholar.org/CorpusID:9814695}
\BIBentrySTDinterwordspacing

\bibitem{hpcorpus_kadosh}
T.~Kadosh, N.~Hasabnis, T.~Mattson, Y.~Pinter, and G.~Oren, ``Quantifying openmp: Statistical insights into usage and adoption,'' 09 2023, pp. 1--7.

\bibitem{monocoder_kadosh}
T.~Kadosh, N.~Hasabnis, V.~A. Vo, N.~Schneider, N.~Krien, M.~Capotă, A.~Wasay, G.~Tamir, T.~Willke, N.~Ahmed, Y.~Pinter, T.~Mattson, and G.~Oren, ``Monocoder: Domain-specific code language model for hpc codes and tasks,'' in \emph{2024 IEEE High Performance Extreme Computing Conference (HPEC)}, 2024, pp. 1--7.

\bibitem{WangFarui2021Atod}
F.~Wang, W.~Zhang, H.~Guo, M.~Hao, G.~Lu, and Z.~Wang, ``\BIBforeignlanguage{eng}{Automatic translation of data parallel programs for heterogeneous parallelism through openmp offloading},'' \emph{\BIBforeignlanguage{eng}{The Journal of supercomputing}}, vol.~77, no.~5, pp. 4957--4987, 2021.

\bibitem{auto_migrate_seq_to_hetero}
C.-Y. Liang, S.-Y. Fu, Y.-P. Liu, and W.-C. Hsu, ``Automatically migrating sequential applications to heterogeneous system architecture,'' in \emph{2018 International Conference on High Performance Computing \& Simulation (HPCS)}, 2018, pp. 114--121.

\bibitem{Bruce2018RN1}
\BIBentryALTinterwordspacing
B.~R. Bruce and J.~Petke, ``Towards automatic generation and insertion of {OpenACC} directives,'' no. RN/18/04, London, UK, 12 Apr. 2018. [Online]. Available: \url{http://www.cs.ucl.ac.uk/fileadmin/UCL-CS/research/Research_Notes/RN_18_04.pdf}
\BIBentrySTDinterwordspacing

\bibitem{chen_ompserial_dataset_2023}
\BIBentryALTinterwordspacing
L.~Chen, Q.~I. Mahmud, H.~Phan, N.~Ahmed, and A.~Jannesari, ``Learning to parallelize with openmp by augmented heterogeneous ast representation,'' in \emph{Proceedings of Machine Learning and Systems}, D.~Song, M.~Carbin, and T.~Chen, Eds., vol.~5.\hskip 1em plus 0.5em minus 0.4em\relax Curan, 2023, pp. 442--456. [Online]. Available: \url{https://proceedings.mlsys.org/paper_files/paper/2023/file/8ee477d6175a03d7098fa23641a2d298-Paper-mlsys2023.pdf}
\BIBentrySTDinterwordspacing

\bibitem{hpc_coder_nichols_2024}
D.~Nichols, A.~Marathe, H.~Menon, T.~Gamblin, and A.~Bhatele, ``Hpc-coder: Modeling parallel programs using large language models,'' in \emph{ISC High Performance 2024 Research Paper Proceedings (39th International Conference)}, 2024, pp. 1--12.

\bibitem{chen_ompgpt_2024}
\BIBentryALTinterwordspacing
L.~Chen, A.~Bhattacharjee, N.~Ahmed, N.~Hasabnis, G.~Oren, V.~Vo, and A.~Jannesari, ``Ompgpt: A generative pre-trained transformer model for\&nbsp;openmp,'' in \emph{Euro-Par 2024: Parallel Processing: 30th European Conference on Parallel and Distributed Processing, Madrid, Spain, August 26–30, 2024, Proceedings, Part I}.\hskip 1em plus 0.5em minus 0.4em\relax Berlin, Heidelberg: Springer-Verlag, 2024, p. 121–134. [Online]. Available: \url{https://doi.org/10.1007/978-3-031-69577-3_9}
\BIBentrySTDinterwordspacing

\bibitem{tree-sitter-documentation}
Tree-Sitter, ``Tree-sitter documentation,'' \url{https://tree-sitter.github.io/tree-sitter/}, accessed: May 28, 2025.

\bibitem{baxter_clone_1998}
I.~Baxter, A.~Yahin, L.~Moura, M.~Sant'Anna, and L.~Bier, ``Clone detection using abstract syntax trees,'' in \emph{Proceedings. International Conference on Software Maintenance (Cat. No. 98CB36272)}, 1998, pp. 368--377.

\bibitem{harel_pragformer_2024}
\BIBentryALTinterwordspacing
R.~Harel, T.~Kadosh, N.~Hasabnis, T.~Mattson, Y.~Pinter, and G.~Oren, ``Pragformer: Data-driven parallel source code classification with transformers,'' \emph{Int. J. Parallel Program.}, vol.~53, no.~1, Oct. 2024. [Online]. Available: \url{https://doi.org/10.1007/s10766-024-00778-9}
\BIBentrySTDinterwordspacing

\bibitem{openacc-best-practices-guide}
{OpenACC Organization}, ``Openacc programming and best practices guide,'' \url{https://www.openacc.org/sites/default/files/inline-files/OpenACC_Programming_Guide_0_0.pdf}, accessed: May 28, 2025.

\bibitem{openai_chat_template}
{OpenAI}, ``Chatml: A chat markup language for training and interacting with large language models,'' \url{https://platform.openai.com/docs/guides/text?api-mode=chat}, accessed: May 28, 2025.

\bibitem{grattafiori2024llama3herdmodels}
\BIBentryALTinterwordspacing
A.~Dubey, A.~Jauhri, A.~Pandey, A.~Kadian, and et~al., ``The llama 3 herd of models,'' \emph{CoRR}, vol. abs/2407.21783, 2024. [Online]. Available: \url{https://doi.org/10.48550/arXiv.2407.21783}
\BIBentrySTDinterwordspacing

\bibitem{chen_evaluating_2021}
\BIBentryALTinterwordspacing
M.~Chen and J.~Tworek, ``Evaluating large language models trained on code,'' \emph{ArXiv}, vol. abs/2107.03374, 2021. [Online]. Available: \url{https://arxiv.org/abs/2107.03374}
\BIBentrySTDinterwordspacing

\bibitem{touvron2023llama}
H.~Touvron, T.~Lavril, and Izacard, ``Llama: Open and efficient foundation language models,'' \emph{arXiv preprint arXiv:2302.13971}, 2023.

\bibitem{zhang2023llamaadapter}
R.~Zhang, J.~Han, A.~Zhou, X.~Hu, S.~Yan, P.~Lu, H.~Li, P.~Gao, and Y.~Qiao, ``Llama-adapter: Efficient fine-tuning of language models with zero-init attention,'' \emph{arXiv preprint arXiv:2303.16199}, 2023.

\bibitem{gao2023llamaadapterv2}
P.~Gao, J.~Han, and Zhang, ``Llama-adapter v2: Parameter-efficient visual instruction model,'' \emph{arXiv preprint arXiv:2304.15010}, 2023.

\bibitem{codellama_rozier_2023}
B.~Rozière, J.~Gehring, and Gloeckle, ``Code llama: Open foundation models for code,'' 08 2023.

\bibitem{peft}
S.~Mangrulkar, S.~Gugger, L.~Debut, Y.~Belkada, S.~Paul, and B.~Bossan, ``Peft: State-of-the-art parameter-efficient fine-tuning methods,'' \url{https://github.com/huggingface/peft}, 2022.

\bibitem{dettmers_qlora_2023}
T.~Dettmers, A.~Pagnoni, A.~Holtzman, and L.~Zettlemoyer, ``Qlora: efficient finetuning of quantized llms,'' in \emph{Proceedings of the 37th International Conference on Neural Information Processing Systems}, ser. NIPS '23.\hskip 1em plus 0.5em minus 0.4em\relax Red Hook, NY, USA: Curran Associates Inc., 2023.

\bibitem{unsloth}
\BIBentryALTinterwordspacing
M.~H. Daniel~Han and U.~team, ``Unsloth,'' 2023. [Online]. Available: \url{http://github.com/unslothai/unsloth}
\BIBentrySTDinterwordspacing

\bibitem{wiki_levenshtein_distance}
{Wikipedia contributors}, ``Levenshtein distance --- {Wikipedia}{,} the free encyclopedia,'' \url{https://en.wikipedia.org/w/index.php?title=Levenshtein_distance\&oldid=1279737312}, 2025, [Online; accessed 13-June-2025].

\bibitem{gemini_pro}
\BIBentryALTinterwordspacing
{Google}, ``{Gemini 2.5 Pro API},'' 2025. [Online]. Available: \url{https://ai.google.dev/gemini-api/docs/models}
\BIBentrySTDinterwordspacing

\end{thebibliography}

\end{document}